\documentstyle[twoside,fleqn,npb,amsmath,epsfig]{article}

\renewcommand{\L}{{\cal L}}

\newcommand{\epsilonbar}{{\overline\epsilon}}

\newcommand{\intx}{\int d^4x}
\newcommand{\VL}{\left( \begin{array}{c}}
\newcommand{\VR}{\end{array} \right)}

\def\GG#1{{\Gamma_{#1}}}
\def\dg#1{\frac{\delta\Gamma}{\delta#1}}

\def\dfrac#1#2{\frac{\delta{#1}}{\delta{#2}}}

\def\pslash#1{{\setbox0=\hbox{$#1$}
  \rlap{\ifdim\wd0>.7em\kern.22\wd0\else\kern.1\wd0\fi /}#1}}

\allowdisplaybreaks[1]
\title{Renormalization of the Minimal Supersymmetric Standard Model}
\author{Wolfgang Hollik,
Elisabeth Kraus,
Markus Roth,
Christian Rupp, 
Klaus Sibold,
\underline{Dominik St{\"o}ckinger}\address{Max-Planck-Institut f\"ur Physik,
M\"unchen; Institut f{\"u}r Theoretische Physik, Universit{\"a}t
Bonn, Karlsruhe, Bern, Leipzig; Deutsches
Elektronen-Synchrotron DESY, Hamburg}\thanks{Talk given by D.S.\ at the RADCOR/Loops
and Legs in Quantum Field Theory 2002, September 8--13, Kloster Banz,
Germany.
{E-mail~address:~\tt dominik@mail.desy.de.}}}
\begin{document}
%\begin{flushright}
%DESY--??--???\\
%{\tt hep-ph/0210016}\\
%\end{flushright}
%\vspace{3ex}
%\begin{center}
%%{\Large\bf Renormalization of the minimal supersymmetric standard model\\}
%\vspace{3ex}
%{\large 
%          Dominik St{\"o}ckinger{\renewcommand{\thefootnote}{\fnsymbol{footnote}}
%\footnote{\parbox[t]{10cm}{
%          dominik@mail.desy.de.}}}}
%  \\[2ex]
%  \parbox{10cm}{\small\center\em
%           Deutsches Elektronen-Synchrotron DESY, 
%  \\            D--22603 Hamburg, Germany
%  }
%\setcounter{footnote}{0}
%\end{center}
%\vspace{2ex}
\begin{abstract}
The renormalization of the Minimal Supersymmetric
Standard Model (MSSM) is presented. We describe symmetry identities
that constitute a framework in which the MSSM is completely
characterized and renormalizability can be proven. Furthermore, we
discuss applications of this framework for the determination of
symmetry-restoring counterterms,  the gauge dependence of $\tan\beta$
and the derivation of non-renormalization theorems.
\end{abstract}

\maketitle

In this talk the renormalization of the Minimal Supersymmetric
Standard Model (MSSM) is presented \cite{MSSM}. A framework is set up
where all counterterms are uniquely determined. This comprises a set
of symmetry identities providing a complete characterization of the
MSSM and a set of on-shell renormalization conditions that forbid
on-shell mixing between different physical fields. In this framework
it has been shown that the MSSM is multiplicatively renormalizable,
infrared finite, and that all on-shell conditions can be satisfied
simultaneously. 

This study is motivated by the fact that no
satisfactory regularization for supersymmetric gauge theories is
known. In particular, dimensional regularization breaks
supersymmetry; hence, 
supersymmetry-restoring counterterms have to be calculated and
added. On the other hand, dimensional reduction is
mathematically inconsistent, and therefore its area of
validity is unclear. In practice, renormalization of the MSSM means
first to check whether the chosen regularization preserves the
symmetries and to add --- if necessary --- symmetry-restoring
counterterms, and second to add the usual symmetric counterterms
(corresponding to field and parameter renormalization) in
order to cancel divergences and satisfy renormalization conditions. 

This raises the deeper question of how to formulate the symmetries of
the MSSM at all on the quantum level and on the level of Green
functions. For the Standard Model \cite{Kraus97}
and general supersymmetric Yang--Mills theories
\cite{MPW96ab,HKS00} the answer is known: the Slavnov--Taylor
and Ward identities provide a complete characterization of the
symmetries. Similar identities should also be formulated
in the MSSM.

The outline of the talk is as follows. First the symmetry identities
of the MSSM are presented. Then we draw important conclusions of these
identities, in particular on the proof of renormalizability and the
practical determination of symmetry-restoring
counterterms. Finally, two applications of the symmetry identities are
discussed, concerning the gauge dependence of the parameter
$\tan\beta$ and a new approach to the non-renormalization theorems in
supersymmetric gauge theories.

The basic symmetries of the (electroweak part of the) MSSM are
spontaneously broken SU(2)$\times$U(1) gauge invariance and softly
broken supersymmetry. Clearly, the basic structure of the symmetry
identities in the MSSM can be obtained by combining the results for
the Standard Model \cite{Kraus97} and for general supersymmetric
Yang--Mills theories \cite{MPW96ab,HKS00}. The main symmetry
content is described by a Slavnov--Taylor identity
\begin{equation}
S(\Gamma)=0,
\end{equation}
where $\Gamma$ denotes the generating functional of the one-particle
irreducible Green functions. It combines all information on gauge
invariance and supersymmetry including the quantum corrections to the
transformations and the commutation relations of the
generators. However, the Slavnov--Taylor identity does not fix the
values of the hypercharges $Y_i$, which however are crucial in order
to fix the electric charges $Q_i = T^3_i+{Y_i}/{2}$ correctly. The
$Y_i$ are fixed by a local Ward identity for the U(1)-symmetry:
\begin{eqnarray}
\partial^\mu\frac{\delta\Gamma}{\delta
V'{}^\mu}=-ig'\sum_{{\rm Fields}\ \varphi_i} \frac{Y_i}{2}\
\varphi_i\frac{\delta\Gamma}{\varphi_i} 
+
\begin{array}{c}{\mbox{gauge-fixing}}\\{\mbox{terms}}
\end{array}
\nonumber
\end{eqnarray}
Along with this local Ward identity, global Ward identities describing
global SU(2)$\times$U(1) and $R$ invariance are formulated. The
symmetry breakings are introduced by using external fields with a 
constant shift. For soft supersymmetry breaking a chiral
supermultiplet $(A,a_\alpha,F_A+v_A)$ with a shift in its highest
component is used, and 
for spontaneous breaking of gauge invariance an
SU(2)$\times$U(1)-multiplet $(\hat\Phi+\hat{\rm v})$ is used. Using
these fields, the symmetry identities take the same form as in the
cases with unbroken symmetries, but when the external fields are set
to their constant values, symmetry breaking is described. 

While the basic structure of the symmetry identities seems obvious,
the difficulty lies in the detailed implementation. In fact, it turns
out that the $R_\xi$-gauge requires that in the MSSM the detailed
structure of the external fields $(\hat\Phi+\hat{\rm v})$
appearing in the symmetry identities must differ from the one in the
literature.

The problem is that the MSSM contains an extended Higgs sector, and
even if CP-conservation is assumed the physical CP-odd Higgs boson
$A^0$ and the charged Higgs bosons $H^\pm$ can mix with unphysical
degrees of freedom:
\begin{eqnarray}
A^0&\leftrightarrow&(G^0, A^\mu_{\rm Long}, Z^\mu_{\rm
Long}),\nonumber\\
H^\pm&\leftrightarrow&(G^\pm, W^\pm{}^\mu_{\rm Long}).
\label{mixings}
\end{eqnarray}
However, in the $R_\xi$-gauge fixing terms only the unphysical fields
appear in the gauge-fixing functions (we restrict ourselves to the
case of the neutral fields for simplicity):
\begin{equation}
F^A=\partial^\mu A_\mu,
\quad F^Z=\partial^\mu Z_\mu+\xi M_Z G^0.
\label{Rxigauge}
\end{equation}
In this form, the gauge fixing would even break global
SU(2)$\times$U(1)-invariance and the U(1)-Ward identity. As proposed
in \cite{Kraus:1995jk,Kraus97}, the U(1)-Ward identity can be restored
by using the external field $(\hat\Phi+\hat{\rm v})$ for writing the
gauge-fixing functions. However, if $(\hat\Phi+\hat{\rm v})$ is chosen
as an SU(2)-doublet like in the Standard Model-case, it turns out that
necessarily $A^0$ and/or $H^\pm$ appear in the gauge fixing. Thus we
are lead to the question which multiplet assignment to choose for
$(\hat\Phi+\hat{\rm v})$. The answer has been found in \cite{MSSM}. 
The multiplet structure of $(\hat\Phi+\hat{\rm v})$ is chosen as the
product of the adjoint and doublet representation of
SU(2)$\times$U(1), and there are two of these 8-component multiplets,
one for each Higgs doublet $H_{1,2}$. Then the gauge-fixing functions
can be written as
\begin{equation}
F^a=\partial^\mu V^a_\mu - 2 {\rm Im}((\hat{\Phi}+\hat{\rm
v})_i^a{}^\dagger H_i), 
\end{equation}
so that $F^a$ transforms in the adjoint representation and is
compatible with the U(1)-Ward identity. At the same time, $\hat{\rm
v}$ has enough components that can be adjusted such that the
$R_\xi$-gauge conditions are reproduced for $\hat\Phi=0$, i.e.\
$F^a|_{\hat\Phi\to0}$ does not contain the fields $A^0$, $H^\pm$, and
the $F^{A,Z}$-components coincide with (\ref{Rxigauge}).

\begin{figure}
\parbox{4.2cm}{ \begin{picture}(3,10)\epsfxsize=3cm
 \put(20,-50){\epsfbox{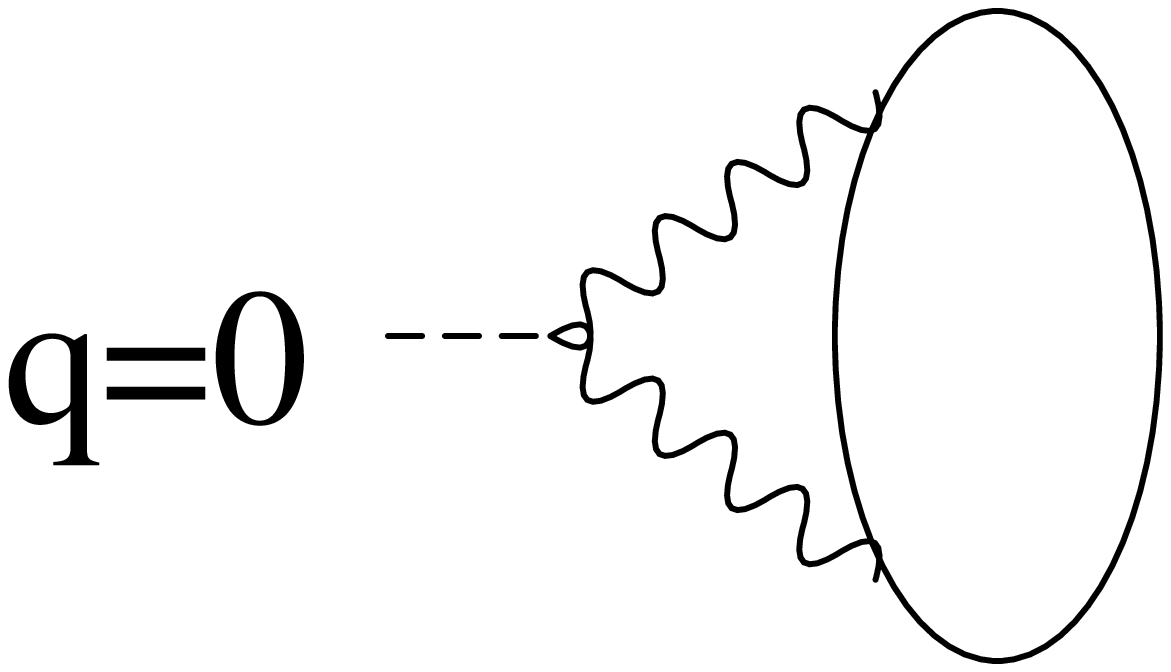}}
 \put(110,-7){$\sim \int\frac{d^4k}{k^4}$}
 \end{picture}}
\caption{Infrared divergences originating from Ward identities
$W\Gamma=\ldots+\intx\hat{\rm v}_i\dg{\hat\Phi_j}$ if counterterms like
$\L_{\rm ct}=\hat{\Phi}_j A^\mu A_\mu$ are present.}
\label{figurediag}
\end{figure}

A more complicated multiplet structure of
this external field, however, also complicates the proof of infrared
finiteness of the MSSM. Consider the counterterm $\L_{\rm
ct}=\hat{\Phi}_j A^\mu A_\mu$ that might be necessary to restore
symmetries as an example. In the calculation of global Ward identities
there appear terms like
$W\Gamma=\ldots+\intx\ \hat{\rm v}_i\dg{\hat\Phi_j}$, leading to
diagrams of the type shown in Fig.\ \ref{figurediag}. Since the
$\hat\Phi_j$-field carries no momentum, all diagrams of this type
contain the infrared divergent integral $\int \frac{d^4k}{k^4}$, no
matter what other external lines and momenta are present. It has to be
shown that such situations cannot arise and that $\Gamma$ itself as
well as all Ward and Slavnov--Taylor identities are infrared finite.
The proof of infrared finiteness can be carried out by identifying the
dangerous components of $\hat\Phi$ and also of $(A,a_\alpha,F_A+v_A)$ using
infrared powercounting, finding an optimal assignment of
the infrared quantum numbers and checking that every dangerous
term is in fact excluded for a symmetry reason. 

Combining the mentioned elements yields the detailed structure of the
symmetry identities in the MSSM. The main identities are the
Slavnov--Taylor and Ward identities in presence of the external
$(\hat\Phi+\hat{\rm v})$ and $(A,a_\alpha,F_A+v_A)$ fields, whose multiplet
structure and infrared powercounting has been determined. There are
further identities corresponding to gauge-fixing conditions, but their
discussion is beyond the scope of the present talk (see sec.\ 3.2 of
\cite{MSSM}). The set of symmetry identities constitutes a full
definition of the MSSM at the quantum level.

Using the symmetry identities we can draw the following conclusions of
basic importance: the symmetry-restoring counterterms are 
uniquely determined, the remaining freedom consists of adding
symmetric counterterms --- which can be shown to correspond to
multiplicative renormalization, and infrared finiteness can be
completely proven.

Furthermore, the Slavnov--Taylor identity can be used to show that all
on-shell renormalization conditions can be satisfied simultaneously
\cite{MSSM}. Since the MSSM is multiplicatively renormalizable, it is
quite obvious that most on-shell conditions can be satisfied be
choosing the field renormalization constants appropriately. An
exception arises from the mixings (\ref{mixings}). The question is
whether it is possible to establish on-shell renormalization
conditions that characterize the fields $A^0$, $H^\pm$ as mass
eigenstates. Using field renormalization it is always possible to
satisfy the on-shell conditions (neglecting the finite widths of the
particles)
\begin{eqnarray}
&&\Gamma_{A^0A^0}(M_A^2)=\Gamma_{A^0G^0}(M_A^2)=0,\nonumber\\
&&\Gamma_{H^+H^-}(M_{H^\pm}^2)=\Gamma_{H^+G^-}(M_{H^\pm}^2)=0.\qquad
\label{OnShell}
\end{eqnarray}
The Slavnov--Taylor identity yields
\begin{eqnarray}
0&=&\sum_{\varphi=A^0,G^0}\GG{c^aY_{\varphi}}\GG{A^0 \varphi}
-\GG{c^a\bar{c}^b}\frac{1}{\xi}\dfrac{F^b}{A^0}\nonumber\\
&&
+\sum_{V=A,Z}\GG{c^aY_{V_\mu}}\GG{A^0V_\mu},
\label{STI}
\end{eqnarray}
and a similar identity for $A^0\to H^\pm$, where $c^a,\bar{c}^a$ are the
Faddeev--Popov (anti-)ghosts and
$Y_\varphi$ denotes the source of the BRS transformation of
$\varphi$. Because of (\ref{OnShell}) and because the $R_\xi$-gauge
can be realized, the first two terms of (\ref{STI}) vanish for
on-shell momenta, and we obtain
\begin{eqnarray}
&&\Gamma_{A^0A^\mu}(M_A^2)=\Gamma_{A^0Z^\mu}(M_A^2)=0,\nonumber\\
&&\Gamma_{H^+W^-{}^\mu}(M_{H^\pm}^2)=0.
\end{eqnarray}
Hence, indeed no on-shell mixing between the physical and unphysical
degrees of freedom occurs and all on-shell conditions can be
satisfied.

This completes the general discussion of 
the renormalization and provides the basis for practical
applications. 

As the first application we will briefly discuss how
supersymmetry-restoring counterterms can be determined in
practice. Generally, the counterterms have to be chosen such that the
symmetry identities hold after renormalization. For supersymmetry, to
kinds of identities are important. Using\footnote{There are fields
$\varphi$ with no corresponding source $Y_\varphi$. For them,
$\dg{\epsilon \delta Y_{\varphi_i}}$ has to be replaced by
$\partial_{\epsilon}s\varphi_i$, where $s\varphi_i$ is the BRS
variation of $\varphi_i$.}
\begin{equation}
0=\dfrac{S(\Gamma)}{\epsilon}=\sum_{{\rm Fields}\ \varphi_i}
\dg{\epsilon \delta Y_{\varphi_i}}\dg{\varphi_i},
\end{equation}
where $\epsilon$ is the ghost corresponding to supersymmetry,
we obtain identities corresponding to supersymmetry relations, where
the prefactors are quantum corrected and renormalized supersymmetry
transformations. Using
\begin{eqnarray}
0&=&\dfrac{S(\Gamma)}{\epsilon\delta\epsilonbar\delta Y_{\varphi_j}}
\nonumber\\
&=&\sum_{{\rm Fields}\ \varphi_i}
\bigg(\dg{\epsilonbar\delta Y_{\varphi_i}}\dg{\epsilon\delta
Y_{\varphi_j}\varphi_i} + (\epsilon\leftrightarrow\epsilonbar)\bigg)
\nonumber\\
&&\pm 2i\sigma^\mu\partial_\mu \varphi_j +\ldots,
\end{eqnarray}
identities corresponding to the supersymmetry algebra are obtained.
Here $\pm$ holds for bosonic/fermionic $\varphi_j$, respectively, and
the dots denote calculable terms corresponding to gauge
transformations and equations of motion in the supersymmetry algebra.
By the requirement that these identities should be satisfied after
renormalization, it is possible to determine first the counterterms
for the renormalized supersymmetry transformations, and then the
counterterms to the vertex functions without BRS insertions. The
results are unique up to symmetric counterterms, which correspond to
multiplicative renormalization. Identities of the first kind have
already been considered in \cite{CJN80}, but they alone do not lead to
unique results for symmetry-restoring counterterms and cannot serve as
tests of the symmetry of a given regularization scheme. Both kinds of
identities have been considered in \cite{SQED,SQCD} for
supersymmetric QED and QCD at the one-loop level. In all cases it has
been found that dimensional reduction preserves the identities at the
regularized level. Thus this scheme is indeed supersymmetric in the
cases considered up to now. 

In the remainder of this talk we want to discuss two applications of
the Slavnov--Taylor identities of the MSSM or other supersymmetric
models: calculating the gauge dependence of the parameter $\tan\beta$
\cite{TB} and deriving non-renormalization theorems
\cite{NRTs,KrausNRT}. Both use certain extensions of the
Slavnov--Taylor identities as tools. The gauge dependence is
calculated using a Slavnov--Taylor identity containing an additional
BRS transformation of the gauge parameter, and the non-renormalization
theorems are derived by introducing BRS transformations of the
coupling constants.

The quantity $\tan\beta$ is one of the main input parameters of the
MSSM. At tree level, 
\begin{equation}
\tan\beta=\frac{v_2}{v_1}.
\end{equation}
 In \cite{TB},
different renormalization schemes for $\tan\beta$ 
were analyzed with the aim to find a scheme that defines $\tan\beta$ at
the same time in a gauge-independent and process-independent way. Using
the extended Slavnov--Taylor identity, the gauge dependence of
$\tan\beta$ can be explicitly calculated for any given scheme. It was
found that in several well-known schemes, the $\overline{DR}$-scheme and
the schemes introduced in \cite{Dabelstein}, $\tan\beta$ is defined in
a gauge-dependent way, i.e.\ the relation between $\tan\beta$ and
observable quantities that can be used to extract $\tan\beta$ from
experiment is gauge dependent. The only exception is the
$\overline{DR}$-scheme, which is gauge independent if its application
is restricted to $R_\xi$-gauges and to the one-loop level.

Hence, a large class of new schemes was considered where $\tan\beta$
is defined in the Higgs sector, as at the tree level. All
gauge-independent schemes in this class were identified. Unfortunately
it turned out that each of these schemes leads to severe numerical
instabilities in the perturbative expansion, so these schemes are not
useful in practice. Given these results, the $\overline{DR}$-scheme
seems to be the best choice of all process-independent definitions for
$\tan\beta$. 

Non-renormalization theorems are among the deepest and most exciting
properties of supersymmetric models. They state the absence of
certain divergences, e.g.\ of quadratic divergences, and thus
provide in particular a solution to the naturalness problem. In
\cite{FlumeKraus}, a new approach towards these theorems has been
developed, which has been applied
to supersymmetric QED and QCD in \cite{NRTs,KrausNRT}. The origin of the
non-renormalization theorems is identified as 
follows. Every term in a supersymmetric Lagrangian is the highest
component of a supermultiplet and thus related to lower-dimensional
field polynomials. Similarly, supersymmetry relates diagrams to
diagrams with a lower degree of divergence.

This fact can be implemented into an extended Slavnov--Taylor identity
by replacing the coupling constants by full supermultiplets. In this
way, every supersymmetric term in the Lagrangian is replaced by a sum
of the form (considering the case of chiral multiplets for simplicity)
\begin{equation}
\label{multiplets}
g\L_{\rm susy} \to g\L_{\rm susy} - \chi\Xi + f A,
\end{equation}
where $(g,\chi,f)$ is the multiplet of the coupling and
$(A,\Xi,\L_{\rm susy})$ the multiplet of the respective term in the
Lagrangian. The higher 
components of the coupling thus couple to the lower components of the
Lagrangian. The supermultiplet structure of the couplings implies BRS
transformations of the couplings. If these are included into the
Slavnov--Taylor identity, an extended identity is 
obtained that can be used to derive non-renormalization theorems. 
%For instance, the relation (\ref{multiplets}) is exploited for Green
%functions by using
%\begin{equation}
%0 = \dfrac{S(\Gamma)}{\epsilon\delta\chi} = 
%\sqrt2\partial_g\Gamma + \dg{\epsilon Y_{\varphi_i}}\dg{\chi\varphi_i}
%.
%\end{equation}

The advantages of this approach are that the non-renormalization
theorems can be derived without assuming a supersymmetric
regularization, in the context of the Wess--Zumino gauge, and that it
makes the underlying algebraic origin apparent. As an illustration, we
list the results for supersymmetric QED. In this model, there are two
independent divergences, conveniently expressed in terms of divergent
renormalization constants for the charge and electron mass:
\begin{eqnarray}
\delta Z_e^{(1)}:\mbox{ only one-loop},\quad
\delta Z_m^{(l)}.
\end{eqnarray}
The renormalization constants
for all soft-breaking parameters can be expressed in terms of $\delta
Z_e$ and $\delta Z_m$:
\begin{eqnarray}
\delta Z_{M_\lambda}^{(1)} & = & 2\delta Z_e^{(1)}:\mbox{ only one-loop},\\
\delta Z_{b}^{(l)} & = & (2l M_\lambda m/b+1)\delta Z_m^{(l)},\\
\delta Z_M^{(l)} & = & \frac12 l(l+1) (M_\lambda^2/b)\delta
Z_m^{(l)}. 
\end{eqnarray}
In addition to relating all renormalization constants, these equations
imply that the charge and the photino mass counterterms are finite
from the two-loop level on and that the two scalar mass counterterms
are only logarithmically divergent.

Similar results can also be derived for non-abelian supersymmetric
gauge theories. In this case a deep connection between the form of the
non-renormalization theorems and two anomalies --- the Adler--Bardeen
anomaly and a supersymmetry anomaly in presence of the supercoupling
\cite{KrausNRT} --- is exhibited. As a byproduct, also the
non-renormalization of the Adler--Bardeen anomaly coefficient can be
proven in a simple way.

To summarize, we started with a discussion of the symmetry identities
of the MSSM. Particular attention was payed to the mixing of physical
and unphysical fields and the implications on the external fields
$(\hat\Phi+\hat{\rm v})$, $(A,a_\alpha,F_A+v_A)$ and their structure. Once
the detailed form of the fields and the symmetry identities was
established, it was possible to prove the 
renormalizability of the MSSM. On the practical side, the
symmetry identities constitute an important tool. We have shown that
they can be used for the unambiguous determination of possible
symmetry-restoring counterterms, for calculating the gauge dependence
of $\tan\beta$, and in a new approach to the non-renormalization
theorems.

%\begin{flushleft}

%\end{flushleft}


\begin{thebibliography}{AA}
\def\ap#1#2#3{{\it Ann. Phys. }{\bf #1~}(19#2)~#3}
\def\app#1#2#3{{\it Acta Phys. Polon. }{\bf #1~}(19#2)~#3}
\def\arnps#1#2#3{{\it Annu. Rev. Nucl. Part. Sci. }{\bf #1~}(19#2)~#3}
\def\fp#1#2#3{{\it Fortschr. Phys. }{\bf #1~}(19#2)~#3}
\def\hepph#1{{\bf hep-ph}/#1}
\def\hepth#1{{\bf hep-th}/#1}
\def\ijmp#1#2#3{{\it Int. Jour. Mod. Phys. }{\bf #1~}(19#2)~#3}
\def\jetp#1#2#3{{\it JETP Lett. }{\bf #1~}(19#2)~#3}
\def\mpl#1#2#3{{\it Mod. Phys. Lett. }{\bf #1~}(19#2)~#3}
\def\npb#1#2#3{{\it Nucl. Phys. }{\bf B #1~}(19#2)~#3}
\def\plbold#1#2#3{{\it Phys. Lett. }{\bf B#1~}(19#2)~#3}
\def\plbnew#1#2#3{{\it Phys. Lett. }{\bf B#1~}(#2)~#3}
\def\prd#1#2#3{{\it Phys. Rev. }{\bf D#1~}(19#2)~#3}
\def\prep#1#2#3{{\it Phys. Rep. }{\bf #1~}(19#2)~#3}
\def\prl#1#2#3{{\it Phys. Rev. Lett. }{\bf #1~}(19#2)~#3}
\def\ptp#1#2#3{{\it Prog. Theor. Phys. }{\bf #1~}(19#2)~#3}
\def\rmp#1#2#3{{\it Rev. Mod. Phys. }{\bf #1~}(19#2)~#3}
\def\rnc#1#2#3{{\it Riv. Nuovo Cim. }{\bf #1~}(19#2)~#3}
\def\sjnp#1#2#3{{\it Sov. J. Nucl. Phys. }{\bf #1~}(19#2)~#3}
\def\zpc#1#2#3{{\it Z. Phys. }{\bf C #1~}(19#2)~#3}
\def\epjc#1#2#3{{\it Eur. Phys. J. }{\bf C #1~}(19#2)~#3} 

\bibitem{MSSM} W.~Hollik, E.~Kraus, M.~Roth, C.~Rupp, K.~Sibold,
D.~St\"ockinger, 
%``Renormalization of the minimal supersymmetric standard model,''
{\it Nucl.\ Phys.}\  {\bf B 639} (2002) 3.
%\bibitem{DReg}   G. 't Hooft, M. Veltman, \npb{44}{72}{189};
%C. Bollini, J. Giambiagi, {\em Nuovo Cim.} {\bf 12B} (1972) 20;
%P. Breitenlohner, D. Maison, {\em Comm. Math. Phys. }{\bf 52}{ (1977) } 11, 39, 55.
%\bibitem{Siegel79} W. Siegel, \plbold{84}{79}{193}.
%\bibitem{Siegel80} W. Siegel, \plbold{94}{80}{37}.
\bibitem{Kraus97} E. Kraus, \ap{262}{98}{155}.
\bibitem{MPW96ab} N. Maggiore, O. Piguet, S. Wolf, \npb{458}{96}{403},
\npb{476}{96}{329}.
\bibitem{HKS00}
W.~Hollik, E.~Kraus, D.~St\"ockinger,
%``Renormalization of supersymmetric Yang-Mills theories with soft  supersymmetry breaking,''
{\it Eur.\ Phys.\ J.\ } {\bf C 23} (2002) 735.
\bibitem{Kraus:1995jk}
E.~Kraus, K.~Sibold,
%``Rigid invariance as derived from BRS invariance: The Abelian Higgs model,''
{\it Z.\ Phys.}\  {\bf C 68} (1995) 331.
%\bibitem{BPHZL}
%\cite{Zimmermann:1969jj}
%\bibitem{Zimmermann:1969jj}
%W.~Zimmermann,
%``Convergence Of Bogoliubov's Method Of Renormalization In Momentum Space,''
%Commun.\ Math.\ Phys.\  {\bf 15} (1969) 208
%[Lect.\ Notes Phys.\  {\bf 558} (1969) 217];
%\cite{Lowenstein:ug}
%\bibitem{Lowenstein:ug}
%J.~H.~Lowenstein,
%``Bphz Renormalization,''
%in {\it C75-08-17.2}
%NYU-TR11-75
%{\it Lectures given at Int. School of Mathematical Physics, Erice, Sicily, Aug 17-31, 1975}.
\bibitem{CJN80} D.M. Capper, D.R.T. Jones, P. van  Nieuvenhuizen, \npb{167}{80}{479}.
\bibitem{SQED} W. Hollik, E. Kraus, D. St{\"o}ckinger, \epjc{11}{99}{365}.
\bibitem{SQCD} W. Hollik, D. St{\"o}ckinger, {\em Eur. Phys. J. }{\bf
C 20} (2001) 105.
\bibitem{TB} A.~Freitas, D.~St\"ockinger,
%``Gauge dependence and renormalization of tan(beta) in the MSSM,''
arXiv: hep-ph/0205281.
\bibitem{NRTs} E.~Kraus, D.~St\"ockinger,
%``Nonrenormalization theorems of supersymmetric QED in the Wess-Zumino  gauge,''
{\it Nucl.\ Phys.}  {\bf B 626} (2002) 73,
%\cite{Kraus:2001ny}
%\bibitem{Kraus:2001ny}
%E.~Kraus and D.~St\"ockinger,
%``Non-renormalization theorems in softly broken SQED and the soft  beta-functions,''
{\it Phys.\ Rev.}\  {\bf D 64} (2001) 115012,
%E.~Kraus and D.~Stockinger,
%``Softly broken supersymmetric Yang-Mills theories: Renormalization and  non-renormalization theorems,''
{\it Phys.\ Rev.}\  {\bf D 65} (2002) 105014.
%%CITATION = HEP-PH 0201247;%%
%%CITATION = HEP-PH 0107061;%%
%\cite{Kraus:2001tg}
%\bibitem{Kraus:2001tg}
\bibitem{KrausNRT} E.~Kraus,
%``An anomalous breaking of supersymmetry in supersymmetric gauge theories  with local coupling,''
{\it Nucl.\ Phys.}\  {\bf B 620} (2002) 55,
%%CITATION = HEP-TH 0107239;%%
%\cite{Kraus:2001id}
%\bibitem{Kraus:2001id}
%E.~Kraus,
%``Calculating the anomalous supersymmetry breaking in Super-Yang-Mills  theories with local coupling,''
{\it Phys.\ Rev.}\  {\bf D 65} (2002) 105003.
%%CITATION = HEP-PH 0110323;%%
%\cite{Kraus:2001kn}
%\bibitem{Kraus:2001kn}
%%CITATION = HEP-TH 0105028;%%
%\cite{Kraus:2002uh}
%\bibitem{Kraus:2002uh}
\bibitem{Dabelstein} A. Dabelstein, \zpc{67}{95}{495},
 P. Chankowski, S. Pokorski, J. Rosiek,
\npb{423}{94}{437}.
\bibitem{FlumeKraus} R.~Flume, E.~Kraus,
%``Non-renormalization theorems without supergraphs: The Wess-Zumino  model,''
{\it Nucl.\ Phys.}\  {\bf B 569} (2000) 625.
\end{thebibliography}
\end{document}